\documentclass[aps,prl,preprint,superscriptaddress]{revtex4}
\usepackage{graphicx}

\begin{document}
\title{Photovoltage Detection of Edge Magnetoplasmon Oscillations and Giant Magnetoplasmon Resonances in A Two-Dimensional Hole System}
\author{Jian Mi}
\affiliation{International Center for Quantum Materials, Peking University, Beijing, 100871, China}

\author{Jianli Wang}
\affiliation{International Center for Quantum Materials, Peking University, Beijing, 100871, China}

\author{L. N. Pfeiffer}
\affiliation{Department of
Electrical Engineering, Princeton University, Princeton, New Jersey
08544, USA}

\author{K. W. West}
\affiliation{Department of
Electrical Engineering, Princeton University, Princeton, New Jersey
08544, USA}

\author{K. W. Baldwin}
\affiliation{Department of
Electrical Engineering, Princeton University, Princeton, New Jersey
08544, USA}

\author{Chi Zhang}
\altaffiliation{Electronic address: gwlzhangchi@pku.edu.cn}
\affiliation{International Center for Quantum Materials, Peking University, Beijing, 100871, China}
\affiliation{Collaborative Innovation Center of Quantum Matter, Beijing, 100871, China}



\begin{abstract}
  In our high mobility p-type AlGaAs/GaAs two-dimensional hole samples, we originally observe the $B$-periodic oscillation induced by microwave (MW) in photovoltage (PV) measurements. In the frequency range of our measurements (5 - 40 GHz), the period ($\Delta B$) is inversely proportional to the microwave frequency ($f$). The distinct oscillations come from the edge magnetoplasmon (EMP) in the high quality heavy hole system. Simultaneously, we observe the giant plasmon resonance signals in our measurements on the shallow two-dimensional hole system (2DHS).
\end{abstract}

\maketitle

In the high quality two dimensional electron system (2DES) in GaAs/AlGaAs heterostructure, many novel electron states and fancy phenomena were revealed by various remarkable tools, especially the microwave techniques.
In transport experiments, different classical and quantum oscillations were discovered continuously.
At a low magnetic field ($B$), $1/B$-periodic microwave induced resistance oscillations (MIRO) and associated zero resistance state (ZRS) were discovered ~\cite{Mani, Zudov, Zudov2001, CLYang}.
Consequently, the edge magnetoplasmon (EMP) oscillations were discovered by photovoltage measurements, and the results indicate that the EMP mode propagates non-locally along the semiconductor heterostructure sample edge ~\cite{Kukushkin2004, Stone}.


The magnetoplasmons (MPs) of 2DES in a perpendicular magnetic field are the hybrid of the classical bulk plasmons and cyclotron resonances, following the dispersion equation:
\begin{equation}
 \label{plasmon}
    \omega_{mp} = \sqrt{\omega_{C}^2+\omega_{p}^2}
  \end{equation}
 , where $\omega_{p}$ is the plasmon frequency related to geometric size of 2DEG and $\omega_{C}= eB/m^{*}$ is the cyclotron resonance frequency with $m^{*}$ the electron effective mass in GaAs ~\cite{Ando, Mast}.
Usually, the spectrum of the 2D (bulk) plasmon can be expressed as: $\omega_{p}^2 = \frac{2\pi n_{s}e^2}{\varepsilon_{eff} m^{*}} |q|$ , and the wave vector $q=\pi /w$ is from the geometric width $w$ ~\cite{Allen1983, KukushkinPRL2003, Mikhailov}.

There also exist the collective chiral excitations propagating at the edge of 2DEG: edge magnetoplasmons (EMPs), have attracted much attention in recent years, for the researches of the quantum Hall effect and the related edge state ~\cite{Mast, Fetter, Heitmann, Wassermeier, Aleiner, Simovic, Mikhailov}.
MPs and EMPs are typically studied by microwave absorption method and time-domain technique.
In microwave assisted transport measurements on high mobility electron gas, a new type of $B$-periodic oscillation is originated from EMPs ~\cite{Kukushkin2004}.
The period is found to follow the relation $\Delta B \propto n_{s}/\omega L$, where $n_{s}$ is the 2D electron density, $\omega$ is the MW-frequency, and $L$ is the distance between measured Hallbar contacts.
Consequently, nonlocal EMPs are revealed by photovoltage measurements ~\cite{Stone}.
By now, only a few results are reported on the MPs in 2DHS, and the EMP oscillations in the hole system have never been observed.

In the past years, a new type of C-doped p-type GaAs/AlGaAs wafers have been developed dramatically, and supply a platform for researching new physics.
A variety of new discoveries were accompanied by the growing mobility in 2D hole system ~\cite{Yuan, TMLuAPL2008, Nichele, ShayeganPRL2011, YLiuPRL2014, CZPRB2015}.
Very recently, some new proposals for EMP are reported, including the topological EMP in the 2DES ~\cite{Jin-Lu}.

In this report, we have studied the MPs and the EMPs by measuring the photovoltage (PV) on 2D heavy hole system in a p-type GaAs/AlGaAs Quantum well (QW).
In contrast to previous research ~\cite{Kukushkin2004, Stone, ManiPRB2013}, we originally observe a giant resonant peak and $B$-periodic oscillations.
Our analysis indicates that the former comes from the bulk plasmon coupled with cyclotron resonance, and the latter are the consequence of the edge magnetoplasmon (EMP) modes.
Similar to those results in 2DES, the period $\Delta B$ is found to decrease with increasing frequency ~\cite{Stone}.
Moreover, the EMP oscillations persist from low to high magnetic fields of 3 - 4 T, which indicates a much weaker damping than those in n-type GaAs/AlGaAs QW ~\cite{Stone}.

Our experiments are performed in a He3 refrigerator, with a base temperature of 300 mK. This project is carried out in a 17.5 nm wide GaAs/Al$_{0.24}$Ga$_{0.76}$As quantum well, with a high density ($p=$2$\times 10^{11}$ cm$^{-2}$) and an ultrahigh mobility ($\mu=2 \times 10^{6}$ cm$^{2}$/Vs) hole system embedded at 130 nm beneath the sample surface. In our measurements, the Hallbar sample with a size of 75 $\mu$m $\times$ 25 $\mu$m is defined by UV-lithography. The high frequency signal is generated by a continuous wave generator and guided down to the base via a semi-rigid coaxial cable, radiating the sample by a linear dipole antenna. The resistance is measured by passing a low-frequency (17 Hz) current $I = 100$ nA through the Hall bar, detecting the voltage drop. The PV measurements are carried out by chopping the microwave radiation at 991 Hz and detecting the voltage drop of two contacts of the Hall bar at the chop frequency using a lock-in amplifier without excitation current. The microwave frequency can range from 5 to 40 GHz.

Figure 1(A) and (B) exhibit an example for photovoltage features under a low temperature of 1.7 K and a microwave (20 GHz) irradiation.
For minimizing the Shubnikov de-Haas (SdH) oscillations at low temperatures (below 1 K), the temperature around 1.7 K is optimal.
Panel (A) give us a general impression on the power-dependent PV results: two giant peaks exist at both positive and negative magnetic fields (for $B \sim \pm 0.3$ T, as marked by the arrows).
The amplitude of the PV signal increases dramatically with the microwave power: for $P = 18$ dBm irradiation from the MW source, the maximum of voltage peak reaches around 200 $\mu$V, surrounded by a saddle-shape envelope.
By zooming in the details above 0.4 T, some oscillations can be clearly illustrated for both positive and negative $B$-field, as shown in Panel (B).
The oscillations are roughly $B$-periodic, and the period of the oscillations keeps constant under varying power.
For the low $B$-field regime ($B: -0.5 \sim +0.5$ T), the oscillations are weakened by the giant resonant signals.

\begin{figure}
\includegraphics[width=0.8\linewidth]{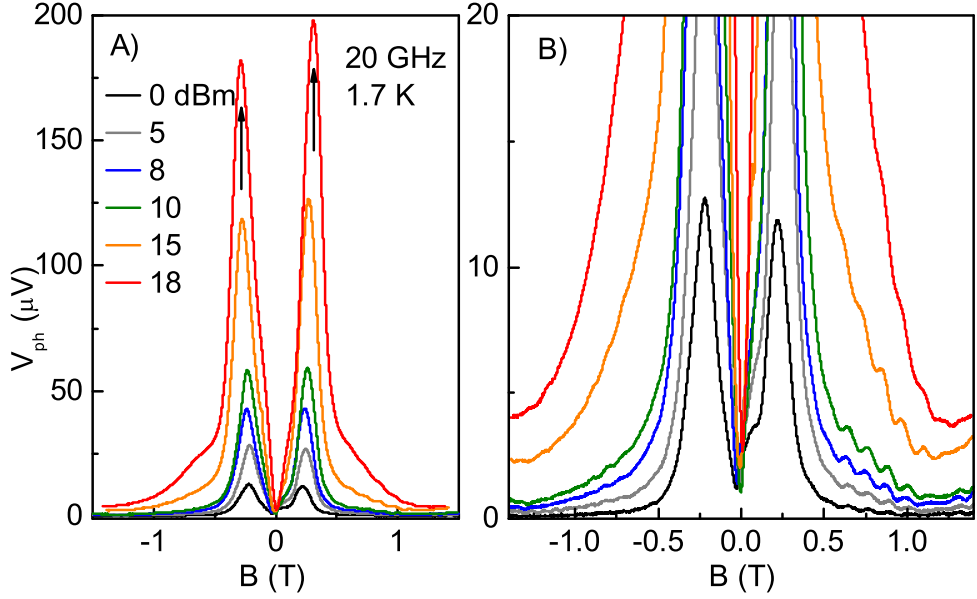}
\caption{(Color online). The figure exhibits the power-dependent photovoltages (PV) $V_{ph}$ at a frequency of 20 GHz. Panel (A): the $PV_{xx}$ is irradiated by a microwave signal of 20 GHz at 1.7 K. Panel (B): the $B$-periodic oscillations of PV are exhibited in the zoom-in scale.}
  \label{FIG1}
 \end{figure}

The photovoltage data of the sample irradiated by microwave signal ranging of 20 - 40 GHz is shown in Fig. 2(A), in which the traces have been offset vertically.
The background photovoltage is lower than 10 $\mu$V. Various contact configurations of our Hallbar sample show the same types of MP resonance and EMP oscillation features.
As the microwave frequency ($f$) increases, the resonance position $B_{R}$ moves to higher $B$. We plot the relation of  $B_{R}$ vs. $f$ in the Fig. 2(B), which is found to approximately follow the relation: $\omega_{mp} = \sqrt{\omega_{C}^2+\omega_{p}^2}$.
In this panel (B), the cyclotron resonance $f=\omega_{C}/2\pi=eB/2\pi m^{*}$ is shown by the blue curve with a heavy effective of $m^{*} \sim 0.6 m_{e}$.
Ideally, the spectrum of 2DHS ($f$ vs. $B$) cross the $B=0$ axis at $f = 36$ GHz ~\cite{Yuan}.
However, for our shallow 2DHS sample, the MP spectrum can be expressed as $\omega_{p}^2=\frac{4\pi p e^{2}q}{[\varepsilon_{eff} coth(qh)+1]\varepsilon_{0}m^{*}}$, where $h$ is the distance from the 2DHS to the surface and the wave vector $q$ can be estimated with the Hallbar width $w=25$ $\mu$m.
In the case of limit $h=0$, the spectrum of the MP could be changed into the cyclotron resonance (the blue curve in FIG. 2(B)).
Under a microwave radiation at 20 GHz, the temperature-dependent PV features are shown in FIG. 2(C).
The giant resonance peak weakens with increasing temperature and disappears at $T$ above $25$ K.
The lineshape of the peak also broadens with temperature, indicating that the phonon scattering plays an important role in the cyclotron resonance damping and broadening.
In Panel (D), the power-dependence results show that the amplitude is proportional to square root of the MW power and the electric component of the microwave (or EM-) field.
Essentially, nominally (nonperfect) Ohmic contacts lead to the non-linearity and non-inhomogeneity.
Thus when the microwave frequency equals the cyclotron frequency, Landau level transitions happen and the band structure at the interface of the contacts varies, leading to the changes in the electronchemical potential and the detected photovoltage ~\cite{ManiPRB2013}.
Consequently, the microwave power helps the enhancements in chemical potential changes and the resonant amplitudes.
Moreover, the resonance position $B_{R}$ changes slightly with MW power.
This may be from the heating effect of microwave power, because the peak positions of $B$-field increases with raising power and temperature (in FIG. 2(C) and (D)).
Microwave heating causes the peak broadening and damping of the resonances.
However, in the temperature range of $T > 1.7$ K, microwave heating/radiating effect is indiscernible with the thermometer in He3 refrigerator.


\begin{figure}
\includegraphics[width=0.8\linewidth]{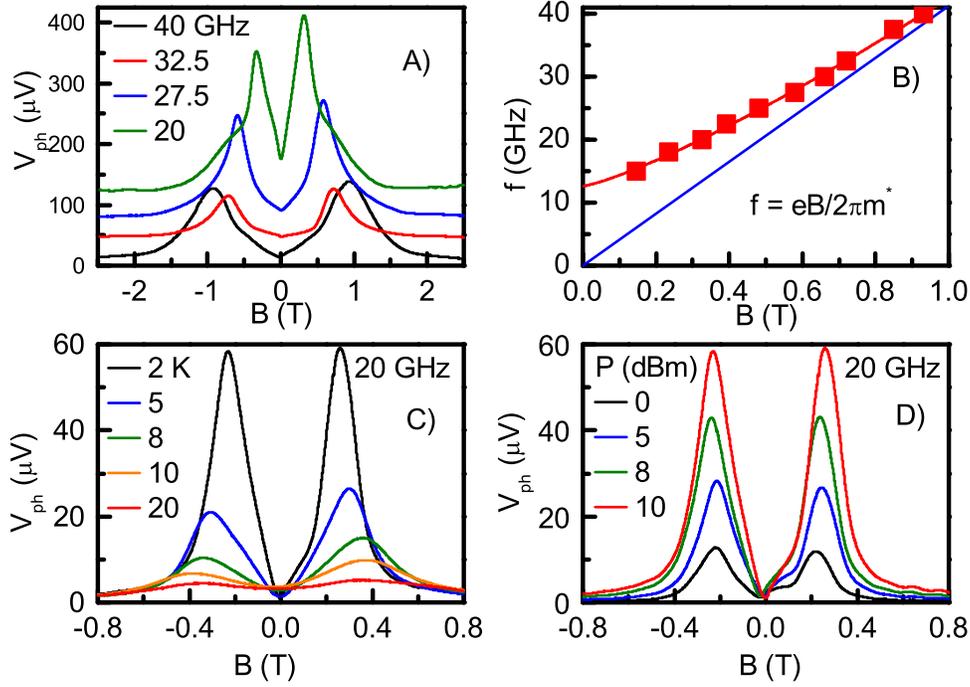}
\caption{(Color online). (A) PV results of the sample under microwave whose frequency ranging from 20 to 40 GHz. (B) High frequency ($f$) vs Resonance position: The red squares and red line are the data measured and the fitting line. The crossing point of the red line with y-axe is the plasmon frequency. The blue line is the cyclotron resonance frequency vs magnetic field theoretically following the equation marked on the insert. (C) Temperature dependence of photovoltage at 20GHz radiation. (D) Power-dependence of PV results at $f = 20$ GHz.}
  \label{FIG2}
 \end{figure}

In the frequency range of 10 - 25 GHz, the $B$-periodic oscillations exist under both positive and negative magnetic fields, as shown in Fig. 3(A) and (B), respectively.
The $B$-field positions of the oscillations of various index $N$ is shown in Panel (C).
Obviously, the oscillation under high magnetic fields is $B$-periodic, and the slope indicates the period of the oscillations.
Our observation is similar to reports of the $B$-periodic EMP oscillations persisting up to 20 K in an ultraclean 2D electron system in GaAs/AlGaAs QW  ~\cite{Kukushkin2004, Stone}.
In the 2DES with an effective mass of 0.067$m_{e}$, the period $\Delta B$ of the EMP oscillations indicate an inverse proportion to the microwave frequency.
Differently, these $B$-periodic oscillations are much weaker for above 3 K, due to the rapid decay of the amplitudes in heavy hole system.
However, the EMP oscillations are universal and robust for $T$ below 3 K.
In Panel (D), the relation between the oscillation period and the high frequency is shown.
The Plot for the $\Delta B$ vs. $1/f$ shows a linear, but not proportional feature, because the intercept is not zero.
In our estimation, the slope is $\sim (3 - 4)$ T $\cdot$ GHz.
A observable intercept can be derived in this figure, $\Delta B \propto (1/f + 1/f_{0})$, here $f_{0} \sim (50 - 60)$ GHz in our measurements.

\begin{figure}
\includegraphics[width=0.8\linewidth]{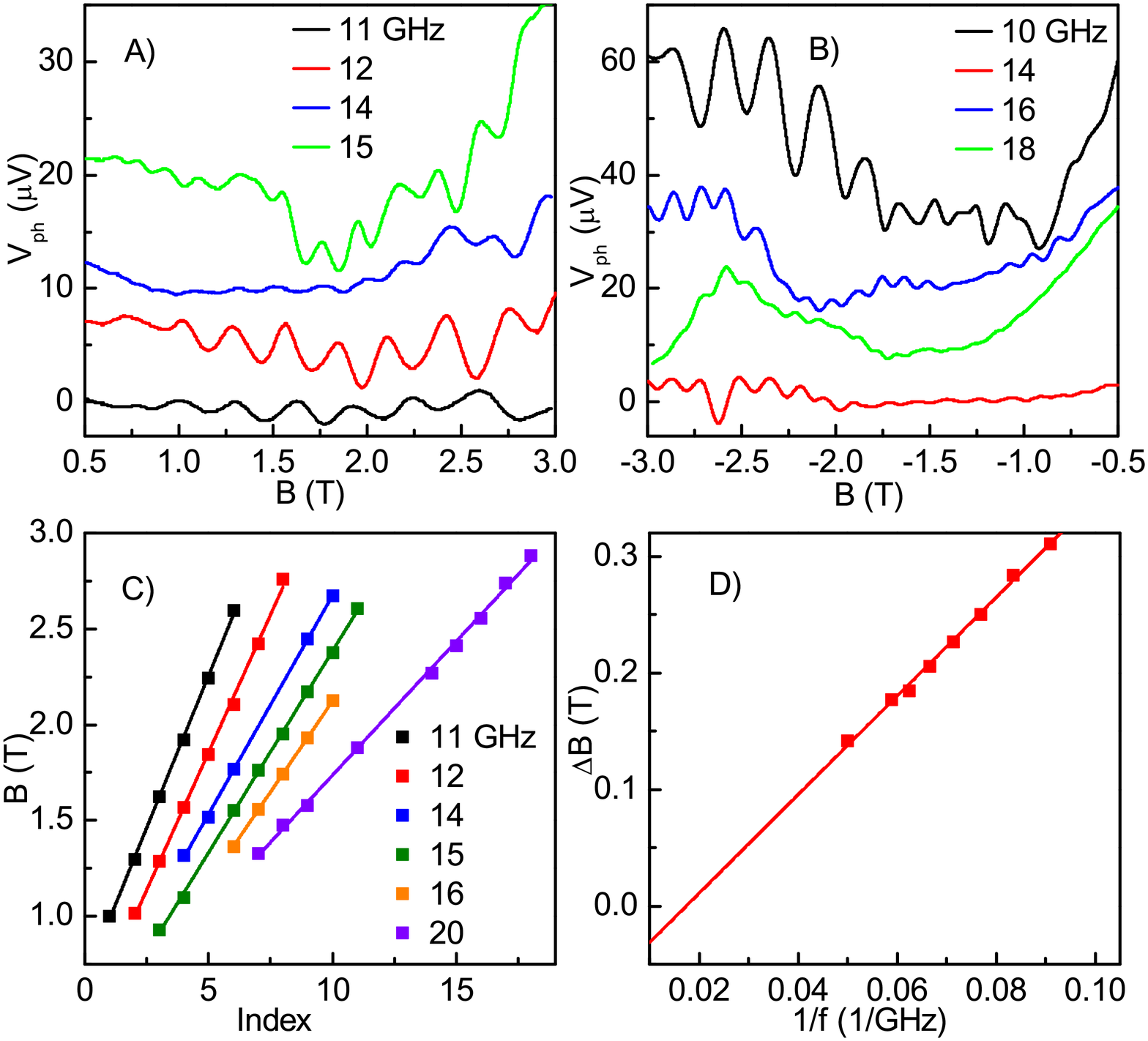}
\caption{(Color online). Panel (A) Photovoltaic oscillations of the contact configuration (PV1) at positive magnetic field for frequency from 11 to 15 GHz.
Panel (B): PV oscillations (PV1) at negative magnetic field for frequency from 10 to 18 GHz.
Panel (C): $B$-positions of the oscillation peaks vs. their index of PV1. The lines here are the linear fitting results, indicating that the PV oscillations are $B$-periodic, and the period can be determined by the slope of each fitting line.
Panel (D): The period $\Delta B$ versus the inverse of high frequency ($1/f$).}
  \label{FIG3}
 \end{figure}

For a simple model of EMP as a narrow charged strip, the spectrum can be expressed as: $\omega_{emp} \propto \sigma_{xy} q$, where $\sigma_{xy} \propto n_{s}/B$ is the Hall conductivity.
The EMP frequency $\omega_{emp} \propto n_{s}N/BL$ comes from the discrete wave vectors $q = 2\pi N/L$, ($N = 1, 2, 3...$).
Thus the expression of periods is simply $\Delta B \propto n_{s}/L f$, where $L$ is the measured contacts distance along the sample edge not the crow-fly distance.
The nonlocal transport property of EMP oscillations is revealed in the PV measurements in an ultraclean 2DES ~\cite{Stone}.
In respect to the results of 2DES, we would like to figure out the transport of EMP mode along the 2D hole sample edge.
Fig. 4(A) compares the EMP oscillations in positive ($B +$) and negative magnetic field ($B -$) for the same contact configuration.
The features are very symmetric with respect to magnetic field, and the oscillation periods are consistent for both positive and negative $B$-fields.
Fig. 4(B) displays the comparison on the photovoltaic oscillations in different contact configurations, which are shown in the inset for Panel (B).
The consistent periods of oscillations in PV1 and PV2 indicate the nonlocal property of the EMP oscillations in our Hallbar sample.

\begin{figure}
\includegraphics[width=0.8\linewidth]{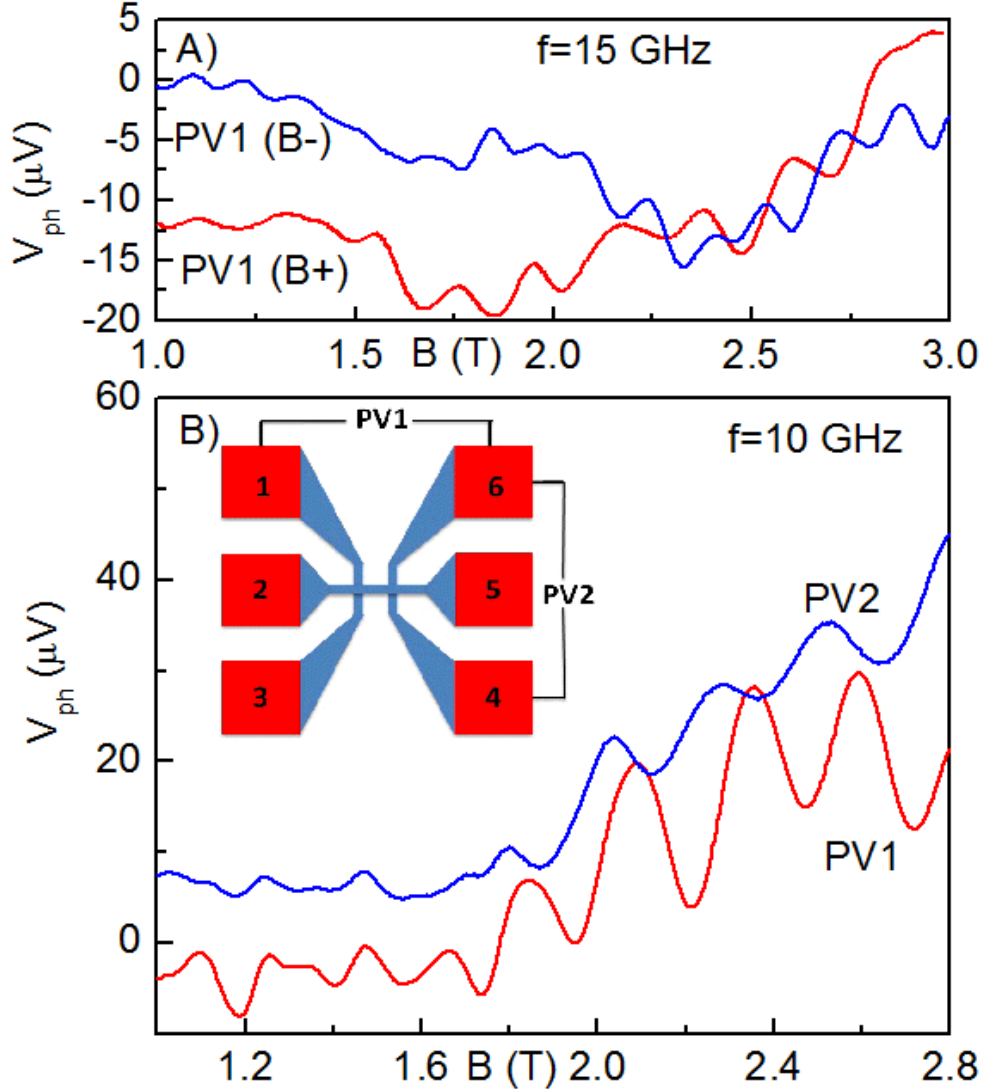}
\caption{(Color online). Panel (A): shows the photovoltage at a frequency of 15 GHz under positive and negative magnetic fields. Panel (B): shows the same $B$-periods of PV oscillations in different configurations for contacts. The inset indicates the schematic of our Hallbar design.}
  \label{FIG4}
 \end{figure}

Based on our discussions, the observation of MP resonance and EMP oscillations depends on the effective mass of the carrier.
In previous experiments, the effective mass were measured in high mobility hole wafers on [001] GaAs orientation substrates ~\cite{Yuan, TMLuAPL2008, Nichele}.
The valence band structure is complex, the heavy hole (HH) and light hole (LH) bands mixture changes the mass.
For the (double-interface) quantum well samples, the hole mass depends on the carrier density and the chemical potential.
An increasing density $p$ moves the $E_{F}$ close to the anticrossing regime in the valence band structure, therefore a larger mass can be observed ~\cite{TMLuAPL2008}.
In the observation of temperature-dependent SdH oscillations, the amplitudes are very sensitive to the excited temperature ~\cite{Nichele}.
Therefore, we estimate the mass of the heavy hole sample in a temperature-dependent measurement (shown in Fig. 5)~\cite{Isihara}.
Below 1 K, the SdH oscillations exhibit large amplitudes.
In the contrary, above 1 K, the SdH amplitudes are weakened distinctly.
Consistently with the results on 15 nm wide QW in Ref. ~\cite{Nichele}, our obtained mass varies with the magnetic fields, with a range of (0.55 $\sim$ 0.65) $m_{e}$ (shown in the inset of Fig. 5) ~\cite{Mass}.
This mass range is consistent with the mass ($\sim 0.6 m_{e}$) from the cyclotron resonance in Fig. 2(B).
Besides, the quantum scattering time $\tau_{q}$ is $\sim 26 - 40$ ps for the SdH oscillation at low temperature, and the transport lifetime $\tau_{tr}$ $\sim 410$ ps is obtained from the sample mobility.

\begin{figure}
\includegraphics[width=0.8\linewidth]{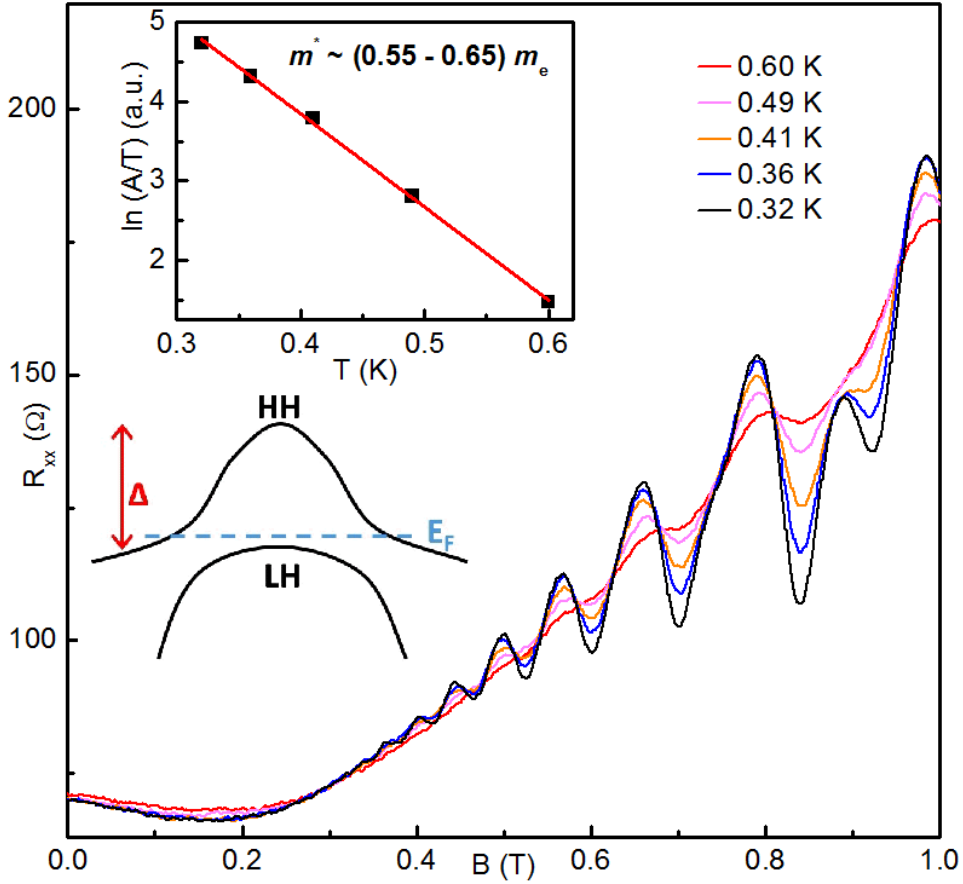}
\caption{(Color online). The 2D hole effective mass can be derived from the temperature-dependent SdH oscillations. The inset on the top shows the range of mass, which can be obtained from the slope of $ln (A/T)$ versus temperature. The inset on the bottom shows the band structure for the high density heavy hole system.}
  \label{FIG4}
 \end{figure}

In our comparison, there are some differences for EMP modes between the n-type and p-type GaAs/AlGaAs QWs.
First of all, in the observations of 2DES, the EMP oscillations coexist with the microwave induced resistance oscillations (MIRO) in the high frequency range ~\cite{Kukushkin2004, Stone}.
However, only clear $B$-periodic oscillations are observed in our measurements.
By now, MIRO has not been detected even in high quality 2DHS sample, including our observations.

Secondly, the EMP oscillations indicate a different spectrum in 2DHS from  those in 2DES ~\cite{Stone}.
With an assumption of carriers with a sharp boundary, the EMP spectrum can be derived as: $\omega_{emp} \propto \sigma_{xy} q ln[\frac{1}{ql(\omega_{emp})}+1]$, and the characteristic length $l(\omega_{emp})$ is identified as the charge stripe width ~\cite{Volkov, SAMikhailov}.
For convenience, complex number form of length is $l(\omega_{emp})=l_{0}+i l_{1}$.
In the low frequency limit ($\omega /\omega_{C} \sim 0$, or $\omega \tau \ll 1$), the length equals the real part: $l(\omega_{emp}) \rightarrow l_{0}$.
In the high-frequency limit ($\omega /\omega_{C} \sim \infty$, or $\omega \tau \gg 1$), the length goes to the imaginary part: $l(\omega_{emp}) \rightarrow l_{1}$ ~\cite{Volkov, SAMikhailov}.
However, our case is neither the low nor high frequency limit due to the ratio $\omega/ \omega_{C} (\propto m^{*})$ is close to 0.2.
Thus the length function is more complex and the simply relation $\Delta B \propto 1/f$ deviates.
A plausible explanation may come from the large effective mass in 2D hole system.
The EMP oscillation spectrum in 2DHS should be more complex than that is 2DES ~\cite{SAMikhailov}.

On the other hand, for temperature below 3 K, the low damping exists in our measurements for high magnetic fields.
In those reports on very high mobility ($\mu \sim 10^{7}$ cm$^{2}$/Vs) 2DES, the EMP oscillations exist up to $\sim 1$ T ~\cite{Stone}.
However, in our high quality hole sample with a mobility of $\mu \sim 2 \times 10^{6}$ cm$^{2}$/Vs, the EMP oscillations persist up to 3 - 4 T.

In summary, in our high mobility p-type GaAs/AlGaAs quantum well, we originally observed giant magnetoplasmon (MP) resonances and edge magnetoplasmon (EMP) oscillations in the photovoltage measurements. Different with the oscillations in 2DES, the periods of EMP oscillations show a different spectrum versus microwave frequencies. Furthermore, it is very striking that the EMP oscillations keep robust at high magnetic fields $B > 3$ Tesla, showing much weaker damping in the heavy hole system. Possibly, it provides a good platform to study the topological edge magnetoplasmon ~\cite{Jin-Lu}.

We would like to thank Rui-Rui Du and Changli Yang for valuable discussions, thank Rui-Rui Du for sample collaboration. This project at Peking University is supported by National Basic Research Program of China (Grant Nos. 2013CB921903 and 2014CB920904) and the National Science Foundation of China (Grant No.11374020). The work at Princeton University is funded by the Gordon and Betty Moore Foundation through the EPiQS initiative Grant GBMF4420, and by the National Science Foundation MRSEC Grant DMR-1420541. J.M. and C.Z. performed experiments; J.M. and C.Z. analyzed data and wrote the paper; J.W. and J.M carried out the cleanroom work; L.P., K.W. and K.B. grew the semiconductor wafers; C.Z. conceived and supervised project.

\end{document}